\include{psfig}
\documentstyle[12pt,twoside,fullpage,epsf]{article}

\title{
\footnote
{We warmly thank P.Rossi, G.Parisi, G.C.Rossi, N.Cabibbo and C.Destri
for interesting discussions and precious suggestions}
Lattice Perturbation Theory by Langevin Dynamics}
\author{
Francesco Di Renzo \\
Giuseppe Marchesini \\
Paolo Marenzoni \\
Enrico Onofri\\
\\
Dipartimento di Fisica\\
Universit\`a di Parma, \\
Viale delle Scienze \\
43100 Parma ( Italy )
}
\date{}

\begin{document}
\maketitle
\thispagestyle{empty}
\vspace{1cm}

\begin{abstract}
We present here an application of the standard Langevin dynamics to
the problem of perturbative expansions on the Lattice QCD. This
method can be applied in the computation of the most general
observables. In this work we will concentrate in particular on the
computation of the perturbative terms of the $1\times 1$ Wilson loop,
up to fourth order. It is shown that a stochastic gauge fixing is
a possible solution to the problem of divergent fluctuations
which affect higher order coefficients.
\end{abstract}

\vskip 1cm
Since its introduction in 1981 by Parisi and Wu \cite{Parisi},
Langevin dynamics has been extensively used for Monte Carlo
simulations.

Basically it consists in a stochastic dynamical system on the field
configuration space dictated by the general equation
\begin{equation}
 \frac{\partial\phi(x,t)}{\partial t} = - \frac{\partial S[\phi]}
 {\partial \phi(x,t)} + \eta(x,t), \
\end{equation}
where $\phi$ is the field, $S[\phi]$ the action and $\eta$ a
Gaussian random noise satisfying to the normalization
\begin{equation}
 <\eta(x,t)\eta(x^{'},t^{'})>=2\delta(x^{'}-x)\delta(t^{'}-t). \
\end{equation}
As a matter of fact, stochastic dynamics is devised in such a way that
time averages on the noise converge to averages on the Gibbs measure
\begin{equation}
 < \frac{1}{T}\int_{0}^{T} dt O[\phi(t)]>_{\eta} \to
 \frac{1}{Z}\int D\phi O[\phi] e^{-S[\phi]}.
\end{equation}

To obtain from equation (1) a useful expression for computer simulations,
one can take $t$ discrete with time step $dt=\epsilon$:
\begin{equation}
 \phi(x,t_{n+1})=\phi(x,t_n) - f_x[\phi,\eta],
\end{equation}
where
\begin{equation}
 f_x=\epsilon  \frac{\partial S}{\partial\phi(x,t_n)}+\sqrt{\epsilon}
 \eta(x,t_n)
\end{equation}
and now $\eta$ is normalized by:
\begin{equation}
 <\eta(x_i,t_i)\eta(x_j,t_j)>=2\delta_{x_i x_j}\delta_{t_i t_j}.
\end{equation}
In this discrete form, Langevin equation has to be regarded solely
as an approximation of equation (1), valid only for $\epsilon\to 0$.

The method has been widely adopted as an alternative to Metropolis,
Heat Bath and other Monte Carlo algorithms for scalar fields and gauge
theories.  In 1985 Batrouni et al. \cite{Bat} presented a new
analysis for lattice fields theories.  Given the standard Wilson
action
\begin{equation}
 S = -\frac{\beta}{2n}\sum_P Tr(U_P+U_P^{\dagger}),
\end{equation}
where the sum is over the plaquettes $P$, in a four dimensional
periodic lattice. $U_P$ are ordered products of the link gauge
variables $U_{\mu}(x)$, which are $SU(3)$ matrices. Here $\mu=1,\ldots
4$ and $x$ is a point of the lattice. Each configuration is then
described by $4\times volume$ $3\times 3$ complex matrices.

For each link $U=U_{\mu}(x)$, one adopts the evolution
\begin{equation}
 U(t_{n+1}) = e^{-F(t_n)} U(t_n), \
\end{equation}
where
\begin{equation}
 F(t_n) = \frac{\epsilon\beta}{4n}\left[\sum_{U_P\supset U_{\mu}}
 (U_P-U_P^{\dagger})-\frac{1}{n}\sum_{U_P\supset U_{\mu}}Tr(U_P-U_P^{\dagger})
 \right]+\sqrt{\epsilon} H(t_n), \\
\end{equation}
and $H$ is a traceless antihermitian noise matrix with normalization
given by
\begin{equation}
 <H_{ik}(x,t)\overline{H}_{lm}(x^{'},t^{'})>_H=[\delta_{il}\delta_{km}-
 \frac{1}{n}\delta_{ik}\delta_{lm}]\delta_{x,x^{'}}\delta_{t,t^{'}}. \
\end{equation}

Langevin approach was originally formulated for perturbation theory
also on the continuum. What we present here is the application of this
idea to compute the weak coupling expansion directly in the lattice.
The problem is well known and has been considered by diagrammatic
technique (see \cite{Pisani}), which allows the calculation of the
expansion coefficients up to $g^4$ (and in some cases $g^6$). Since
gauge fields are written as
\begin{equation}
\begin{array}{lcr}
 U_{\mu}=e^{gA_{\mu}},&A_{\mu}^{\dagger}=-A_{\mu},&TrA_{\mu}=0\\
\end{array}
\end{equation}
where $g$ is the coupling constant, the Langevin equation takes the
following form:
\begin{equation}
 e^{gA^{'}_{\mu}}=e^{-F}e^{gA_{\mu}}.
\end{equation}
The fields $A_{\mu}$ can be expanded in series of $g$
\begin{equation}
 A_{\mu}=\sum_{k}g^k A_{\mu}^{(k)}.
\end{equation}
In the same manner, recalling that $\beta=1/g^2$ and imposing
$\epsilon=g^2\tau$, the drift (9) becomes;
\begin{equation}
 F_{\mu}=\frac{\tau}{12}\left[\sum_{U_P\supset U_{\mu}}
 (U_P-U_P^{\dagger})-\frac{1}{3}\sum_{U_P\supset U_{\mu}}Tr(U_P-U_P^{\dagger})
 \right]+g\sqrt{\tau} H(t_n)=\sum_{k}g^kF^{(k)}, \\
\end{equation}
The main point is to apply to equation (12) the Baker - Campbell -
Hausdorff formula and to extract the contributions order by order in
$g$.

At present, we have implemented the simulation computing the
evolutions for the gauge fields $A_{\mu}$ up to fourth order in $g$.
At this order, perturbative coefficients of many observables have been
computed analytically. Thus, in order to check our lattice formulation
of the above Langevin dynamics, the terms of the standard $1\times 1$
plaquette have been measured (always to the order $g^4$), confirming
the analytical results.

While the original motivation of the Langevin approach was to
make it possible to calculate in perturbation theory without
fixing a gauge, it is known that some divergent fluctuations
(averaging to zero) may plague high order terms. We confirm
this phenomenon in the case of the plaquette expansion.
We observe indeed that, from the third order in $g$, the errors
associated to our observables grow in time,
 even if the mean value remains always
stable around its known value. In higher orders,
this spurious fluctuation may completely hide the
signal.
As a way out we have applied a technique which goes
back to Zwanziger and has been more recently implemented on the
lattice \cite{Pietro} in the form of ``stochastic gauge fixing''.
We may think to a new source in the Langevin equation, responsible of
a stochastic gauge fixing, so that the actual
algorithm implemented is (for the gauge links)
\begin{equation}
\begin{array}{c}
 U_{\mu}^{'N}(x)=e^{F[U_{\mu}^{N},x,\mu]}U_{\mu}^{N}(x)\\
 \\
 U_{\mu}^{N+1}(x)=e^{w[U_{\mu}^{'N},x]}U_{\mu}^{'N}(x)
 e^{-w[U_{\mu}^{'N},x-\mu]}.\\
\end{array}
\end{equation}
with
\begin{equation}
\begin{array}{c}
 w[U_{\mu},x]=\alpha\sum_{\mu}\left( \Delta_{-\mu}\left[ U_{\mu}(x)-
 U_{\mu}^{\dagger}(x-\mu)\right]\right)_{traceless}\\
\\
 \Delta_{-\mu} U_{\nu}(x) \equiv U_{\nu}(x)-U_{\nu}(x-\mu)
\end{array}
\end{equation}
$\alpha$ being a free parameter. \\ As a matter of fact, the result is
very impressive. We report, for example, the term in $g^4$ of the
plaquette, measured (in figure 1) without gauge fixing, and (in
figure 2) with the above gauge fixing.  By means of this essential
reduction of the noise, the result we obtained for the reported term,
with $\tau$ and $\alpha$ set to 0.02, is $c_4=1.211\pm 0.003$,
through an average over 2048 iterations.

All the numerical experiments have been done on Connection Machine
CM2, using CM Fortran, with a great contribution of the CMIS
assembler.  This kind of computation is indeed very expensive: 2.455
Gflops are needed to complete a single Langevin iteration on a lattice
size of $8^4$, at the fourth order, with the measure of the
corresponding coefficient of the plaquette. For the same lattice size,
the program uses about 10 MBytes of memory.

The work is in progress to go to higher orders and test a wide class
of observables of interest for Lattice gauge Theories.

\begin{figure}[h]
{
        \centerline{\psfig{file=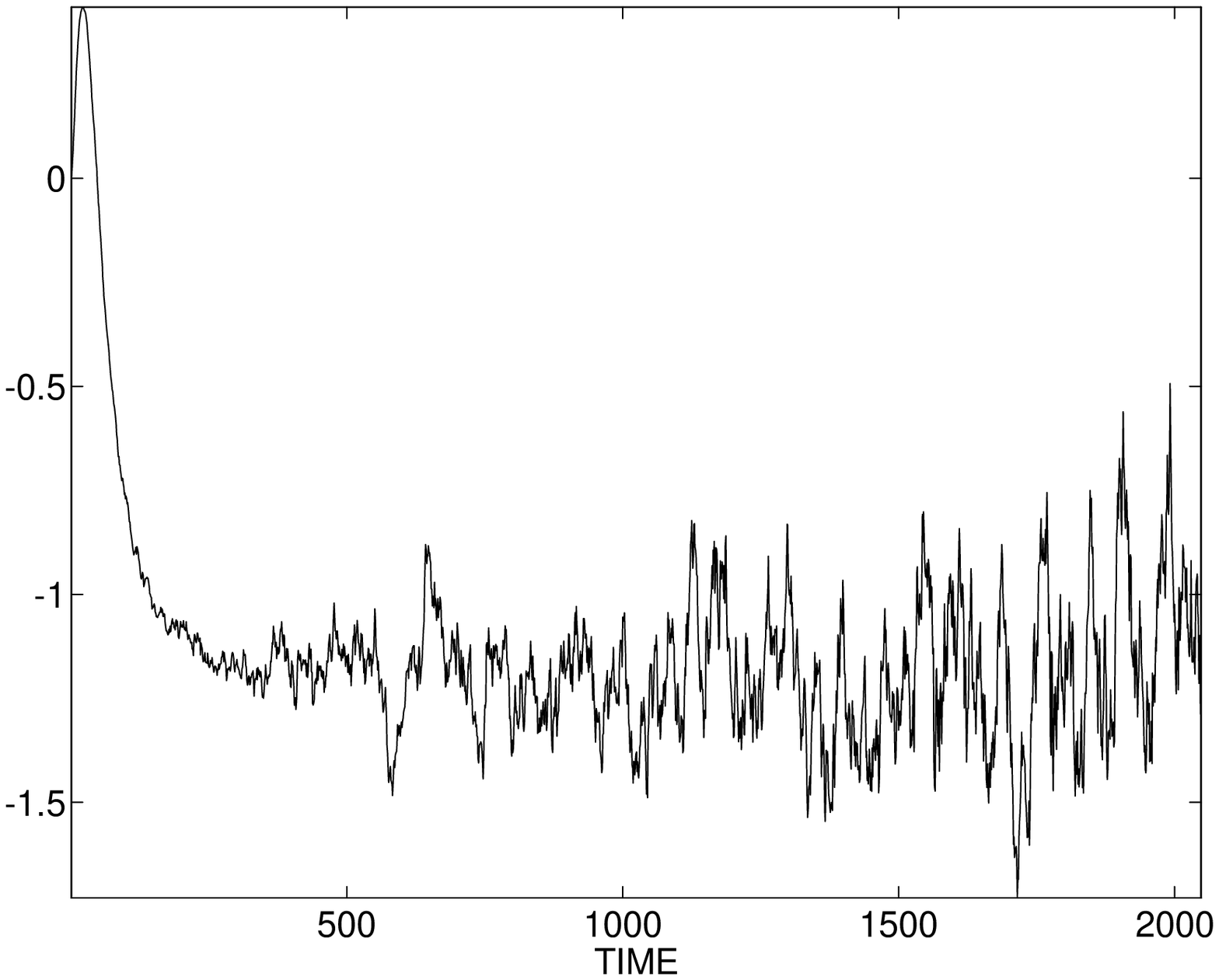}}
}
\caption{Term in $g^4$ of the plaquette, measured without gauge fixing}
\end{figure}
\newpage
\begin{figure}[h]
{
        \centerline{\psfig{file=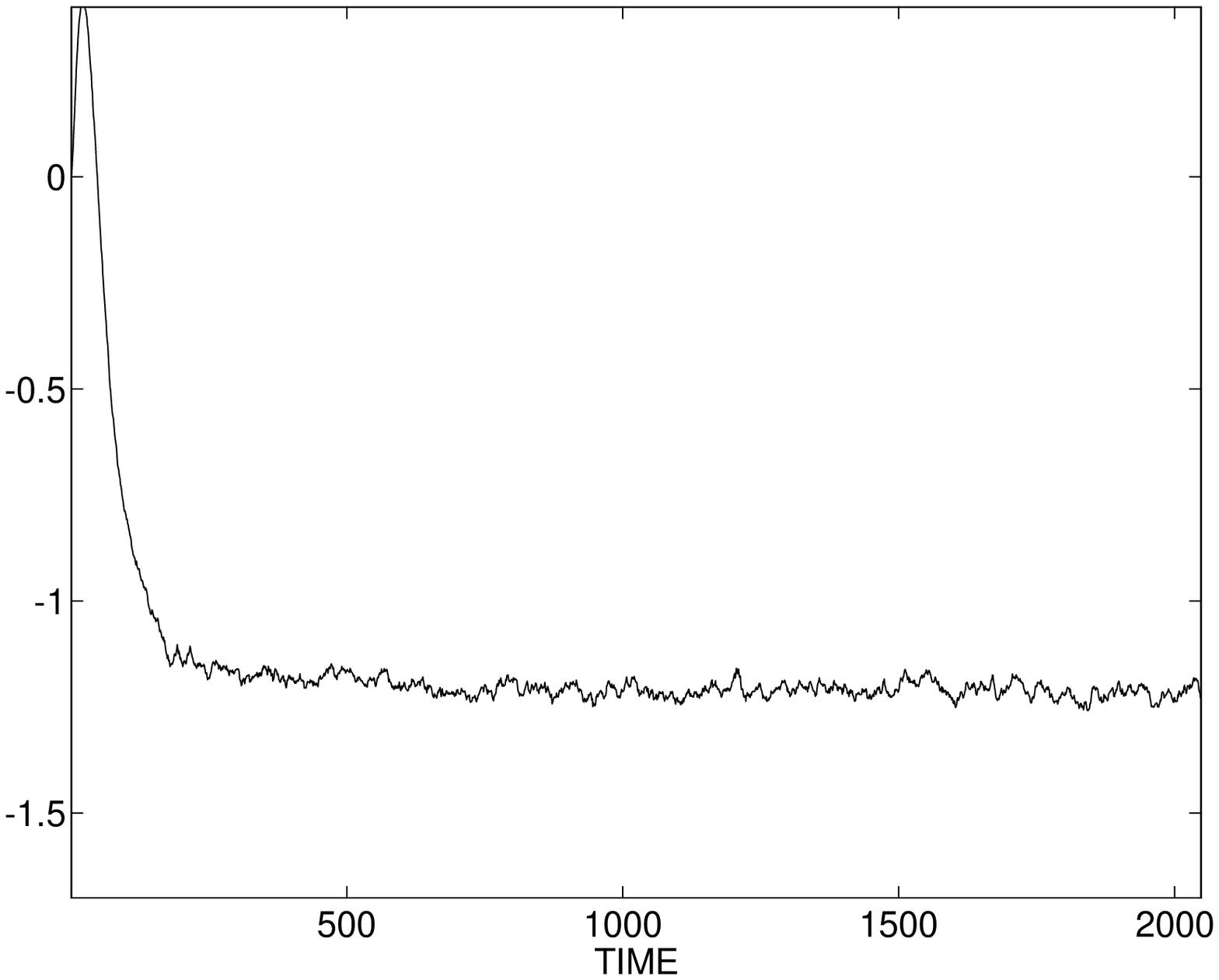}}
}
\caption{Term in $g^4$ of the plaquette, measured with gauge fixing}
\end{figure}

\end{document}